\documentclass[12pt]{article}
\usepackage{amsmath,amsfonts}
\usepackage[nosort]{cite}

\makeatletter\@addtoreset{equation}{section}\makeatother

\def\bR {\mathbb{R}}

\input epsf

\newcommand{\vev}[1]{{\left< {#1} \right>}}

\newcommand{\Tr}{{\rm Tr\,}}

\newcommand{\cD}{{\cal D}}

\newcommand{\cN}{{\cal N}}
\newcommand{\cP}{{\cal P}}
\newcommand{\cS}{{\cal S}}

\newcommand{\preprint}[1]{\begin{table}[t]  %%
             \begin{flushright}               %%
             {#1}                             %%
             \end{flushright}                 %%
             \end{table}}                     %%
\renewcommand{\title}[1]{\vbox{\center\LARGE{#1}}\vspace{5mm}}
\renewcommand{\author}[1]{\vbox{\center#1}\vspace{5mm}}
\newcommand{\address}[1]{\vbox{\center\em#1}}
\newcommand{\email}[1]{\vbox{\center\tt#1}\vspace{5mm}}

\begin{document}

\begin{titlepage}
\preprint{hep-th/0605151 \\ 
}

\title{$1/4$ BPS circular loops, unstable world-sheet instantons 
and the matrix model}

\author{Nadav Drukker}

\address{The Niels Bohr Institute, Copenhagen University\\
Blegdamsvej 17, DK-2100 Copenhagen, Denmark}

\email{drukker@nbi.dk}

\abstract{
The standard prescription for computing Wilson loops in the $AdS$/CFT 
correspondence in the large coupling regime and tree-level involves 
minimizing the string action. In many cases the action has more than one 
saddle point as in the simple example studied in this paper, where there are 
two $1/4$ 
BPS string solutions, one a minimum and the other not. Like in the case of 
the regular circular loop the perturbative expansion seems to be captured 
by a free matrix model. This gives enough analytic control to extrapolate 
from weak to strong coupling and find both saddle points in the 
asymptotic expansion of the matrix model. The calculation also suggests 
a new BMN-like limit for nearly BPS Wilson loop operators.
}

\end{titlepage}

\section{Introduction}

This note deals with a family of circular Wilson loop operators 
in $\cN=4$ supersymmetric 
Yang-Mills theory which have certain couplings to three of the six scalar 
fields. They may be written as
\begin{equation}
W_{\theta_0}=\frac{1}{N}\Tr \cP
\exp\left[\int \left(iA_\alpha\dot x^\alpha(\tau)
+|\dot x(\tau)|\Theta^I(\tau)\Phi_I\right)d\tau\right]\,,
\end{equation}
where $0<\tau<2\pi$, the path in space $x^\alpha(\tau)$ is a circle of 
radius $R$
\begin{equation}
x^1=R\cos \tau\,,\qquad
x^2=R\sin \tau\,,
\label{x1-x2}
\end{equation}
and they couple to the three scalars $\Phi_1$, $\Phi_2$ and $\Phi_3$ in 
the following way
\begin{equation}
\Theta^1=\sin\theta_0\cos\tau\,,\qquad
\Theta^2=\sin\theta_0\sin\tau\,,\qquad
\Theta^3=\cos\theta_0\,,
\end{equation}
with an arbitrary fixed $\theta_0$.

These operators were first presented in \cite{Drukker:2005cu} 
where they were evaluated using the $AdS$/CFT correspondence 
\cite{Maldacena:1997re, Rey,Maldacena-wl} by a classical string 
surface. But these operators are quite interesting and deserves more attention. 
As we shall see they preserve $1/4$ of the supersymmetries of the 
vacuum and by varying $\theta_0$ it is possible to interpolate from 
the usual circle \cite{Berenstein:1999ij,Drukker:1999zq} 
(at $\theta_0=0$) that couples to only one of the scalars and 
preserves half of the supersymmetries, to the $1/4$ BPS circular loop 
\cite{Zarembo:2002an} for $\theta_0=\pi/2$ which couples to 
two scalars.

Evaluating these Wilson loops in $AdS$ requires finding a surface 
which partially wraps an $S^2\subset S^5$. There are two ways to 
do that, over the northern and southern poles, and the resulting values 
of the classical action are $\pm\sqrt{\lambda}\,\cos\theta_0$ 
where $\lambda=g_{YM}^2N$ is the 't Hooft coupling. 
On the gauge theory side evaluating this operator in perturbation theory 
is very similar to the usual circle at $\theta_0=0$ and at least to 2-loop 
order is captured by a Gaussian matrix model 
\cite{Erickson:2000af,Drukker:2000rr}. The only modification is the 
replacement of 
$\lambda\to\lambda'=\lambda\cos^2\theta_0$, hence the 
standard agreement of the string calculation with the matrix model 
result carries over.

What is very intriguing about the $AdS$ calculation is that there is more 
than one saddle point. In general one expects the string expansion to be 
asymptotic, and indeed an extra exponentially suppressed saddle point is 
found contributing $\exp{-\sqrt{\lambda'}}$ to the expectation 
value of the Wilson loop. Up to now only the dominant saddle point 
was considered and was indeed found also from the matrix model. 
But at the planar level the matrix model is given by a Bessel function 
whose asymptotic expansion at large argument has two saddle points, 
which matches the $AdS$ calculation including the subleading one! 
The rest of the paper contains the details of this remarkable agreement.

\section{Gauge theory calculation}

Let us start by calculating those circular Wilson loop observables in 
perturbation theory. At order $g_{YM}^2$ one should sum over the gauge 
field and scalar exchange
\begin{equation}
\begin{aligned}
\vev{W}&=1+
\frac{1}{2N}\int d\tau_1\,d\tau_2\,\Tr T^aT^b\Big[
-\dot x^\alpha(\tau_1)\dot x^\beta(\tau_2)
G_{\alpha\beta}^{ab}(x(\tau_1),x(\tau_2))
\\&\hskip1.2in
+|\dot x^\alpha(\tau_1)||\dot x^\beta(\tau_2)|
\Theta^I(\tau_1)\Theta^I(\tau_2)
G^{ab}(x(\tau_1),x(\tau_2))\Big]\,.
\label{pert-circle1}
\end{aligned}
\end{equation}
Here $T^a$ are generators of the gauge group and they satisfy 
$T^aT^a=N/2\times I$, where $I$ is the identity matrix. 
$G_{\alpha\beta}^{ab}$ is the gauge propagator and 
$G^{ab}$ the scalar propagator, and in the Feynman gauge they 
are given by
\begin{equation}
G_{\alpha\beta}^{ab}(x_1,x_2)
=\frac{g_{YM}^2\delta_{\alpha\beta}\delta^{ab}}{(x_1-x_2)^2}\,,
\qquad
G^{ab}(x_1,x_2)
=\frac{g_{YM}^2\delta^{ab}}{(x_1-x_2)^2}\,.
\end{equation}

Therefore at order $g_{YM}^2$ the circle is given by
\begin{equation}
\vev{W}=1+\frac{g_{YM}^2N}{(4\pi)^2}\int d\tau_1\, d\tau_2
\frac{-\cos(\tau_1-\tau_2)
+\sin^2\theta_0\cos(\tau_1-\tau_2)+\cos^2\theta_0}
{4\sin^2(\tau_1-\tau_2)/2}\,.
\end{equation}
Remarkably, the integrand is a constant, $\frac{1}{2}\cos^2\theta_0$, 
as in the case studied in \cite{Erickson:2000af}, which is recovered by
taking $\theta_0=0$. Thus one finds
\begin{equation}
\vev{W}=1+\frac{g_{YM}^2N}{8}\cos^2\theta_0+O(g^4)\,.
\label{circle-pert}
\end{equation}

In the case where $\theta_0=0$ it was shown \cite{Erickson:2000af} 
that the interacting graphs in the Feynman gauge at order $g^4$ vanish. 
The same also happens at $\theta_0=\pi/2$, where the expectation 
value of the Wilson loop is unity. Quite remarkably this extends to 
arbitrary $\theta_0$. It is possible to separate the graphs to 
contributions associated to $\Phi_3$ and those associated to 
$\Phi_1$ and $\Phi_2$. The graphs involving $\Phi_3$ will be 
identical (up to a constant) to those for $\theta_0=0$ and vanish 
by the same calculation of \cite{Erickson:2000af}. The graphs 
involving the other two scalars are the same as for $\theta_0=\pi/2$ 
and will vanish as well.

So the contribution at order $g_{YM}^4$ comes only from non-interacting 
graphs, where the propagator is a constant. It is then reasonable to conjecture 
that at higher orders interacting graphs still don't contribute and the 
full result will be given by the sum of ladders, as in the case of 
$\theta_0=0$ \cite{Erickson:2000af,Drukker:2000rr}.

Under that assumption the Wilson loop will be given by the sum of all 
non-interacting diagrams which is easily written in terms of a 
0-dimensional Hermitian Gaussian matrix model. The $\theta_0$ 
dependence will show up in the coupling constant, where replacing 
$\lambda\to\lambda'=\lambda\cos^2\theta_0$ will give 
the correct normalization of the matrix propagator
\begin{equation}
\vev{W_{\theta_0}}_\text{ladder}
=\vev{\frac{1}{N}\Tr \exp(M)}_\text{0d}
=\frac{1}{Z}\int \cD M\,\frac{1}{N}\Tr \exp(M)
\exp\left(-\frac{2N}{\lambda'}\Tr M^2\right)\,.
\end{equation}

The full large $N$ expansion of this matrix model was given in 
\cite{Drukker:2000rr}. The leading result at large $\lambda$ is 
\cite{Erickson:2000af}
\begin{equation}
\vev{W_{\theta_0}}_\text{ladder}
\sim\exp\sqrt\lambda\,|\cos\theta_0|
=\exp\sqrt{\lambda'}\,.
\end{equation}

\subsection{Supersymmetry}

The vacuum of $\cN=4$ super Yang-Mills has 32 supersymmetries which 
are generated by the spinors
\begin{equation}
\epsilon (x)=\epsilon_0+\sigma_\alpha x^\alpha \epsilon_1\,,
\end{equation}
with constant $\epsilon_0$ and $\epsilon_1$. Here $\sigma_\alpha$ 
are the usual Dirac matrices and the equations below will also involve 
$\rho^i$, which act on the $SO(6)$ indices of $\epsilon$. The 
conventions used are those of dimensionally reduced $\cN=1$ super 
Yang-Mills in ten dimensions, so all the gamma matrices $\sigma^\alpha$ 
and $\rho^i$ anti-commute.

At the linear order the supersymmetry variation of the Wilson loop is 
proportional to
\begin{equation}
\Big[-i\sigma_1\sin\tau+i\sigma_2\cos\tau
+\sin\theta_0(\rho_1\cos\tau+\rho_2\sin\tau)
+\rho_3\cos\theta_0\Big]\epsilon (x)\,,
\end{equation}
A Wilson loop will be supersymmetric if the above 
expression vanishes for some of the components of $\epsilon(x)$, 
which in this case is
\begin{equation}
\epsilon(x)=\epsilon_0
+R(\sigma_1\cos\tau+\sigma_2\sin\tau)\epsilon_1\,.
\end{equation}
The resulting equation may be separated into terms with different 
functional dependence on $\tau$ which 
should all vanish independently
\begin{equation}
\begin{array}{llrl}
\cos\tau:&\quad&
(i\sigma_2+\sin\theta_0\,\rho_1)\epsilon_0&=
-R \cos\theta_0\,\rho_3\sigma_1 \epsilon_1\,,\\
\sin\tau:&&
(-i\sigma_1+\sin\theta_0\,\rho_2)\epsilon_0
&=-R\cos\theta_0\,\rho_3\sigma_2 \epsilon_1\,,\\
\cos\tau\sin\tau:&&
0&=R\sin \theta_0(\rho_2\sigma_1+\rho_1\sigma_2)\epsilon_1\,,\\
\cos^2\tau:&&
0&=R\sin \theta_0(\rho_1\sigma_1-\rho_2\sigma_2)\epsilon_1\,,\\
1:&&
\cos\theta_0\,\rho_3\epsilon_0&=
R(i\sigma_1-\sin \theta_0\,\rho_2)\sigma_2\epsilon_1\,.
\end{array}
\end{equation}
These conditions are not independent. First for $\sin\theta_0=0$ they are 
all solved as long as $\epsilon_0$ and $\epsilon_1$ are related by
\begin{equation}
\epsilon_0=iR\rho_3\sigma_1\sigma_2\epsilon_1\,.
\end{equation}
This configurations preserves $1/2$ the supersymmetries, but they all 
involve some of the super-conformal transformations.

For $\sin\theta_0=1$ the two constant spinors are not related, rather there 
are two independent conditions on each
\begin{equation}
(i\sigma_2+\rho_1)\epsilon_a=(i\sigma_1-\rho_2)\epsilon_a=0\,.
\end{equation}
Thus this solutions preserves $1/4$ of the regular supersymmetries and 
$1/4$ of the super-conformal ones.

For generic $\theta_0$ it is easy to see that all the above equations are 
satisfied as long as the following two relations hold
\begin{equation}
\begin{gathered}
\cos\theta_0\epsilon_0=
R(-i\sigma_1+\sin \theta_0\,\rho_2)\rho_3\sigma_2\epsilon_1\,,\\
(\rho_2\sigma_1+\rho_1\sigma_2)\epsilon_1=0\,.
\end{gathered}
\end{equation}
Note that as a consequence the last equation holds also for $\epsilon_0$. 
The Wilson loop that is studied here preserves, therefore $1/4$ of the 
supersymmetries.

\section{String theory calculation}

The description of those Wilson loops by strings in $AdS_5\times S^5$ 
was presented in \cite{Drukker:2005cu} (section 4.3.2), the following 
is a short review (in a different coordinate system). Use the target space 
metric
\begin{eqnarray}
ds^2&=&L^2
\left(-\cosh^2\rho\,dt^2+d\rho^2+\sinh^2\rho(d\chi^2
+\cos^2\chi\,d\psi^2+\sin^2\chi\,d\varphi_1^2\right)
\nonumber\\
&&+L^2\left(d\theta^2+\sin^2\theta d\phi^2
+\cos^2\theta\left(d\vartheta_1^2+\sin^2\vartheta_1\left(
d\vartheta_2^2+\sin^2\vartheta_2 d\varphi_2^2
\right)\right)\right)\,.
\label{circle-metric}
\end{eqnarray}
This is global Lorentzian $AdS_5$ with curvature radius $L$ related to 
the 't Hooft coupling by $L^4=\lambda\alpha'^2$. The circle will 
follow the coordinate $\psi$ on the equator of the $S^3$ on the 
boundary of $AdS_5$. On the $S^5$ side the string will be inside an 
$S^2$ given by $\theta$ and $\phi$ (after doubling the range of 
$\theta$ to $[0,\pi]$ at the expense of $\vartheta_1$). Using the ansatz
\begin{equation}
\rho=\rho(\sigma)\,,\qquad
\psi(\tau)=\tau\,,\qquad
\theta=\theta(\sigma)\,,\qquad
\phi(\tau)=\tau\,,\qquad
t=\chi=\vartheta_1=0\,,
\end{equation}
leads to the action in conformal gauge
\begin{equation}
\cS=\frac{L^2}{4\pi\alpha'}
\int d\sigma\,d\tau \left[\rho'^2+\sinh^2\rho
+\theta'^2+\sin^2\theta \right]\,.
\end{equation}

The equations of motion are
\begin{equation}
\begin{aligned}
\rho''&=\sinh\rho\cosh\rho\,,\\
\theta'' &= \sin\theta\cos\theta\,,
\end{aligned}
\end{equation}
and the Virasoro constraint reads
\begin{equation}
\rho'^2+\theta'^2=\sinh^2\rho+\sin^2 \theta\,.
\end{equation}
The first integral for $\rho$ is
\begin{equation}
\rho'^2-\sinh^2\rho=c\,,
\end{equation}
and to get a surface that corresponds to a single circle and not the correlator 
or two one has to set $c=0$, so the solution is
\begin{equation}
\sinh\rho(\sigma)=\frac{1}{\sinh\sigma}\,.
\label{rho-solution}
\end{equation}
An integration constant in this equation that shifts $\sigma$ was set to 
zero so the boundary of the world-sheet at $\sigma=0$ is at the boundary 
of $AdS_5$. Then the first integral for $\theta$ is
\begin{equation}
\theta'^2=\sin^2\theta\,,
\end{equation}
which is solved by
\begin{equation}
\sin\theta(\sigma)=\frac{1}{\cosh(\sigma_0\pm\sigma)}\,,\qquad
\cos\theta(\sigma)=\tanh(\sigma_0\pm\sigma)\,.
\label{theta-solution}
\end{equation}
In this equation the integration constant $\sigma_0$ is important, it is 
fixed by the boundary condition that at $\sigma=0$
\begin{equation}
\cos\theta_0=\tanh\sigma_0\,.
\end{equation}
Depending on the sign in (\ref{theta-solution}) the surface 
extends over the north or south pole of $S^5$.

The bulk part of the classical action is proportional to the area
\begin{equation}
\begin{aligned}
S_\text{bulk}&=\sqrt\lambda\int d\sigma 
\left(\sinh^2\rho+\sin^2 \theta\right)
=\int d\sigma \left(\frac {1}{\sinh^2\sigma}+
\frac{1}{\cosh^2 (\sigma_0\pm\sigma)}\right)
\\
&=\sqrt\lambda
\left(\coth\sigma_\text{min}\mp\tanh\sigma_0\right)
=\sqrt\lambda
\left(\cosh\rho_\text{max}\mp\cos\theta_0\right)\,.
\end{aligned}
\end{equation}
Here $\sigma_\text{min}$ is a cutoff on $\sigma$ and 
$\rho_\text{max}$ the corresponding cutoff on $\rho$. 
This divergent part is canceled by an extra boundary term in the 
action \cite{Drukker:1999zq}, so the final result is
\begin{equation}
\cS=\mp\cos\theta_0\sqrt\lambda\,.
\label{action-circle}
\end{equation}
The two signs correspond to a string extended over the north or south 
poles of $S^2$. From this one finds that the expectation value of the Wilson 
loop at strong coupling is given by
\begin{equation}
\vev{W}\sim \exp\left[
\pm\cos\theta_0\sqrt\lambda\right]\,,
\end{equation}
and the sign should be chosen to minimize the action.

\subsection{Supersymmetry}

In order to check supersymmetry choose the vielbeins (only for the directions 
that are turned on)
\begin{equation}
e^1=L\,d\rho\,,\qquad
e^3=L\sinh\rho\,d\psi\,,\qquad
e^5=L\,d\theta\,,\qquad
e^6=L\sin\theta\,d\phi\,.
\end{equation}

$\Gamma_a$ will be ten real constant gamma matrices and define 
$\gamma_\mu=e_\mu^a\Gamma_a$ and 
$\Gamma_\star=\Gamma^0\Gamma^1\Gamma^2\Gamma^3\Gamma^4$ 
the product of all the gamma matrices in the $AdS_5$ directions. With 
this the dependence of the Killing spinors on the relevant coordinates 
may be written as (see for example 
\cite{Lu:1998nu,Skenderis:2002vf,Kim:2002tj,Canoura:2005uz})
\begin{equation}
\epsilon=e^{-\frac{i}{2}\rho\,\Gamma_\star\Gamma_1}
e^{\frac{1}{2}\psi\,\Gamma_{13}}
e^{-\frac{i}{2}\theta\,\Gamma_\star\Gamma_5}
e^{\frac{1}{2}\phi\,\Gamma_{56}}\epsilon_0\,,
\end{equation}
where $\epsilon_0$ is a chiral complex 16-component spinor. This 
satisfies the Killing spinor equation%
\footnote{
$D_\mu=\partial_\mu+\frac{1}{4}\omega_\mu^{ab}\Gamma_{ab}$
and the only relevant non-zero components of the spin-connection
are $\omega_\psi^{13}=-\cosh\rho$ and 
$\omega_\phi^{56}=-\cos\theta$.}
\begin{equation}
\left(D_\mu+\frac{i}{2L}\Gamma_\star\gamma_\mu\right)
\epsilon=0\,.
\end{equation}

The projector associated with a fundamental string in type IIB is
\begin{equation}
\Gamma=\frac{1}{\sqrt{g}}\,
\partial_\tau x^\mu\partial_\sigma x^\nu
\gamma_\mu\gamma_\nu K\,,
\end{equation}
where $g$ is the induced metric on the world-sheet and $K$ acts on 
spinors by complex conjugation. The number of supersymmetries 
preserved by the string is the number of independent solutions to the 
equation $\Gamma\epsilon=\epsilon$.

For the two solutions (the signs correspond to the two choices in 
(\ref{theta-solution}))
\begin{equation}
\Gamma=\frac{1}{\sinh^2\rho+\sin^2\theta}
\left(\sinh^2\rho\,\Gamma_{13}\pm\sin^2\theta\,\Gamma_{56}
+\sinh\rho\sin\theta\,\Gamma_{16}
\pm\sinh\rho\sin\theta\,\Gamma_{53}\right)K\,.
\end{equation}
The equation has to hold for all $\sigma$ and $\tau$. Since 
$\Gamma_{13}$ commutes with $\Gamma_\star\Gamma_5$ and 
with $\Gamma_{56}$ and also $\Gamma_\star\Gamma_1$ commutes with 
$\Gamma_\star\Gamma_5$ one may write the Killing spinor as
\begin{equation}
\epsilon=e^{-\frac{i}{2}\rho\,\Gamma_\star\Gamma_1
-\frac{i}{2}\theta\,\Gamma_\star\Gamma_5}
e^{\frac{1}{2}\tau(\Gamma_{13}+\Gamma_{56})}\epsilon_0\,.
\label{Killing-2}
\end{equation}
Note that $\Gamma$ does not depend on $\tau$, the only place $\tau$ appears 
in the projector equation is in the second exponential of this expression for 
the Killing spinors. To eliminate this dependence impose the condition
\begin{equation}
(\Gamma_{13}+\Gamma_{56})\epsilon_0=0\,.
\label{SUSY-cond-1}
\end{equation}

Now commuting the terms in the projector $\Gamma$ through the 
remaining exponential in (\ref{Killing-2}), remembering that 
$K$ acts by complex conjugation, one gets
\begin{equation}
\begin{aligned}
\Gamma\epsilon&=\frac{1}{\sinh^2\rho+\sin^2\theta}
\Big[e^{-\frac{i}{2}\rho\,\Gamma_\star\Gamma_1
+\frac{i}{2}\theta\,\Gamma_\star\Gamma_5}
\left(\sinh^2\rho\,\Gamma_{13}
\pm\sinh\rho\sin\theta\,\Gamma_{53}\right)
\\&\hskip1.4in
+e^{\frac{i}{2}\rho\,\Gamma_\star\Gamma_1
-\frac{i}{2}\theta\,\Gamma_\star\Gamma_5}
\left(\pm\sin^2\theta\,\Gamma_{56}
+\sinh\rho\sin\theta\,\Gamma_{16}\right)\Big]K\epsilon_0\,.
\end{aligned}
\end{equation}
Next note that by factoring out by the remaining exponential in the 
Killing spinor, the projector equation $\Gamma\epsilon=\epsilon$ 
reduces to an equation on the constant spinor 
$\bar\Gamma\epsilon_0=\epsilon_0$ with
\begin{equation}
\begin{aligned}
\bar\Gamma&=\frac{1}{\sinh^2\rho+\sin^2\theta}\,
\Big[
e^{i\theta\,\Gamma_\star\Gamma_5}
\left(\sinh^2\rho\,\Gamma_{13}
\pm\sinh\rho\sin\theta\,\Gamma_{53}\right)
\\&\hskip1.4in
+e^{i\rho\,\Gamma_\star\Gamma_1}
\left(\pm\sin^2\theta\,\Gamma_{56}
+\sinh\rho\sin\theta\,\Gamma_{16}\right)\Big]K
\end{aligned}
\end{equation}
By expanding the exponentials and using (\ref{SUSY-cond-1}) 
the projector equation becomes
\begin{equation}
\begin{aligned}
\bar\Gamma\epsilon_0
&=\frac{1}{\sinh^2\rho+\sin^2\theta}\,
\Big[
(\cos\theta\sinh^2\rho\mp\cosh\rho\sin^2\theta)\,\Gamma_{13}
\\&\hskip1.4in
+(\cos\theta\pm\cosh\rho)\sin\theta\sinh\rho\,\Gamma_{16}
\Big]K\epsilon_0\,.
\end{aligned}
\end{equation}
Replacing the solutions for $\rho(\sigma)$ (\ref{rho-solution}) 
and $\theta(\sigma)$ (\ref{theta-solution})
\begin{equation}
\bar\Gamma\epsilon_0
=\frac{1}{\cosh\sigma_0}
\Big[\sinh\sigma_0\,\Gamma_{13}+\Gamma_{16}\Big]K\epsilon_0
=\Big[\cos\theta_0\,\Gamma_{13}
+\sin\theta_0\Gamma_{16}\Big]K\epsilon_0\,.
\end{equation}
Half the eigenvalues of this matrix (for generic $\sigma_0$) are one 
and it commutes with $\Gamma_{1356}$, so the two projections are 
compatible and the string solution preserves $1/4$ of the supersymmetries.

Note that the final expression does not depend on whether the surface 
extends over the north or south pole of the $S^2$ and depends only on 
$\theta_0$. Therefore both those surfaces preserve the same 
supersymmetries which are the same as those found on the gauge 
theory side.

\section{Discussion}

The family of Wilson loop operators considered in this note preserve 
$1/4$ of the supersymmetries of the vacuum and allow to do some 
very interesting calculations. On the gauge theory side it was easy 
to calculate them to order $g_{YM}^4$ and the result 
is the same as for the $1/2$-BPS loop \cite{Erickson:2000af} with 
the replacement $\lambda\to\lambda'=\lambda\cos^2\theta_0$. It 
is therefore natural to conjecture that the final result is given by the same 
matrix model as in \cite{Drukker:2000rr} with this replacement.

On the string theory side two classical string solutions describing this 
Wilson loop were found and both preserved the same supersymmetry. 
For generic $\theta_0$ the action of those two surfaces is not equal, 
rather they have the opposite signs. It is quite common to find more 
than one saddle point to the string equations of motion 
\cite{Drukker:2005cu} and there are indeed more solutions here 
one gets by adding extra wrappings of the sphere, but those would 
not be supersymmetric.

These non-supersymmetric saddle points of the string action will not 
contribute to the same expectation values as the supersymmetric ones 
due to having extra fermionic zero modes. So if the two surfaces found 
here are indeed the only supersymmetric world-sheets satisfying the 
correct boundary conditions, the Wilson loop expectation value will 
have a semiclassical expansion at large $\lambda$ as
\begin{equation}
\vev{W_{\theta_0}}
\sim e^{\sqrt{\lambda'}}+e^{-\sqrt{\lambda'}}\,.
\end{equation}

Recall that at the planar level%
\footnote{For some results beyond the planar level see
\cite{Drukker:2005kx,Yamaguchi:2006tq}} 
the Gaussian matrix model is given by Wigner's semi-circular distribution
\begin{equation}
\vev{W_{\theta_0}}_\text{ladder}
=\frac{2}{\pi\lambda'}
\int_{-\sqrt{\lambda'}}^{\sqrt{\lambda'}}dx\sqrt{\lambda'-x^2}\,e^x
=\frac{2}{\sqrt{\lambda'}}I_1\left(\sqrt{\lambda'}\right)\,,
\label{bessel}
\end{equation}
where $I_1$ is the modified Bessel function. 
Using the asymptotic expansion of the Bessel function at large 
$\lambda'$ 
one finds 
the result that should be reproduced by semiclassical supergravity
\begin{equation}
\vev{W_{\theta_0}}_\text{ladder}
=\sqrt\frac{2}{\pi}\frac{e^{\sqrt{\lambda'}}}{{\lambda'}^{3/4}}
\sum_{k=0}^\infty \left(
\frac{-1}{2\sqrt{\lambda'}}\right)^k
\frac{\Gamma(\frac{3}{2}+k)}{\Gamma(\frac{3}{2}-k)}
-i\sqrt\frac{2}{\pi}\frac{e^{-\sqrt{\lambda'}}}{{\lambda'}^{3/4}}
\sum_{k=0}^\infty \left(\frac{1}{2\sqrt{\lambda'}}\right)^k
\frac{\Gamma(\frac{3}{2}+k)}{\Gamma(\frac{3}{2}-k)}\,.
\label{asymptotic}
\end{equation}
This expression represents two saddle points with classical action 
$\pm\sqrt{\lambda'}$, exactly as was found here from string theory%
\footnote{Those saddle points come from the semiclassical 
evaluation of the integral in (\ref{bessel}) around the two endpoints.}.
The asymptotic expansion includes an infinite series of perturbative 
corrections in inverse powers of $\sqrt{\lambda'}$ which should be 
found by doing the world sheet perturbation expansion around those 
solutions as was pursued in \cite{Drukker:2000ep}.

Note that on $S^5$ there are three transverse directions to the $S^2$. 
Around the minimum of the action those directions are massive, but 
not around the other saddle point. Turning on those deformations of 
the surface will cause it to ``slip'' away from that pole. Those three 
modes are tachyonic and each contributes a factor of $i$ to the fluctuation 
determinant. This too is matched by the results of the asymptotic 
expansion, where the second term in (\ref{asymptotic}) is imaginary.

Clearly the term with negative exponent will never dominate the action 
and its contribution is smaller than any of the $(\lambda')^{-k/2}$ 
corrections to the leading term. It's quite miraculous that it was possible 
to fit this term between the perturbative gauge theory calculation 
and string theory. Such results are often associated with localization 
theorems, which may be the case here due to the large number of 
supersymmetries preserved by this Wilson loop.

One can go much further than the semiclassical string calculation. 
It is possible to take $\lambda$ large, so the string theory is still on a 
low curvature 
background while keeping $\lambda'=\cos^2\theta_0\lambda$ small, 
in a fashion similar to the BMN limit \cite{Berenstein:2002jq}. For 
$\cos\theta_0=0$ Zarembo's solution \cite{Zarembo:2002an} has 
three zero modes parameterizing an $S^3$ with measure
\begin{equation}
d\Omega_3=\frac{1}{2\pi^2}\,d\alpha\sin^2\alpha\,d\Omega_2\,,
\end{equation}
where the range of $\alpha$ is $[0,\pi]$ and $\Omega_2$ is the measure 
on an $S^2$ that remains unbroken for nonzero $\cos\theta_0$. Turning 
on $\cos\theta_0$ leads to a potential
 $\cos\alpha\cos\theta_0\sqrt\lambda$, 
so the integration over the broken zero modes gives for the Wilson loop
\begin{equation}
\vev{W_{\theta_0}}=
\frac{2}{\pi}\int_0^\pi d\alpha\,\sin^2\alpha\,
e^{-\cos\alpha\,\sqrt{\lambda'}}\,.
\end{equation}
This is exactly equal to the result of the matrix model at the planar 
level (\ref{bessel}). Here it is reproduced from perturbative string 
theory by the inclusion only of the zero modes. It would be interesting 
to see if this kind of BMN-like limit generalizes to other deformations 
of supersymmetric Wilson loops.

Recently the $AdS$ description of the supersymmetric Wilson loops 
of the type constructed by Zarembo \cite{Zarembo:2002an} was 
found \cite{Dymarsky:2006ve}. Those operators preserve the regular 
Poincar\'e supersymmetries and have trivial expectation values 
\cite{Guralnik:2003di,Guralnik:2004yc}, but there should be many 
more supersymmetric Wilson loops which preserve 
other combinations of the regular and conformal 
supersymmetry generators. Those include the ones studied in this 
paper as well as some deformations of the line or circle by insertion 
of local operators as in \cite{Drukker:2006xg}. There is 
a very rich structure of supersymmetric Wilson loops that is worth 
exploring.

Clearly some of the usual intuitions about supersymmetry does not 
apply to those combinations of regular and conformal supersymmetries. 
Here there is an unstable surface that preserves some supersymmetry.

In the case of the $1/2$ BPS circle with $\theta_0=0$ it is possible 
to map it to the line which preserves a regular supersymmetry and 
whose expectation value is trivial. For other values of $\theta_0$ 
the resulting line operator will not be trivial, rather it will be 
given by the path
\begin{equation}
x^1=\tau\,,
\end{equation}
for $\tau\in\bR$ and the coupling to the scalars is given by%
\footnote{This is different from the periodic couplings along the line 
considered in \cite{Tseytlin:2002tr,Drukker:2005cu}, which were 
not supersymmetric}
\begin{equation}
\Theta^1=\sin\theta_0\frac{\tau^2-1}{\tau^2+1}\,,\qquad
\Theta^2=\sin\theta_0\frac{2\tau}{\tau^2+1}\,,\qquad
\Theta^3=\cos\theta_0\,.
\end{equation}

It is possible to study this operator both in $AdS$ and at weak coupling. 
By doing a conformal map on the $AdS$ solution one finds a new 
surface with action%
\begin{equation}
\cS=\sqrt\lambda(1\mp\cos\theta_0)\,.
\end{equation}
The difference between the circle (\ref{action-circle}) and this case is 
$\sqrt\lambda$ in agreement with a general argument 
\cite{Drukker:2000rr} that whenever a conformal transformation 
maps a compact Wilson loop to a non-compact one the ratio of the 
Wilson loop VEVs is universal. This ratio is fixed to $\exp\sqrt\lambda$ 
(or more generally to the result of the matrix model) by considering the 
example of $\theta_0=0$.

On the gauge theory side at one-loop the expectation value of the line is
\begin{equation}
\vev{W}=1-\frac{g_{YM}^2N}{8}\sin^2\theta_0\,,
\end{equation}
which also agrees with this general argument. Note that unless 
$\theta_0=0$, the supersymmetries preserved by this line all involve 
the superconformal generators, and it is impossible to find a map that 
will turn them into operators annihilated purely by combinations of 
the Poincar\'e supersymmetry generators.

Another recent development in the study of Wilson loops is the renewed 
interest in the description of them in terms of D-branes 
\cite{Rey,Drukker:2005kx,Yamaguchi:2006tq,Gomis:2006sb,
Okuyama:2006jc,Hartnoll:2006is}. It would be very interesting to find 
the relevant solutions for the $1/4$ BPS Wilson loops, see if there are 
two solutions also for the DBI action, and study the $1/N$ corrections 
to the two saddle points.

\section*{Acknowledgments}
I am grateful to Bartomeu Fiol for collaboration on the early stages of this 
work. I would also like to thank Niklas Beisert, Juan Maldacena, 
Jan Plefka and Gordon Semenoff for useful discussions and correspondences. 
This project was started during a visit to the Weizmann institute, whose 
hospitality is greatly appreciated. Part of the work was also done while 
visiting the Aspen Center for physics which provided an inspiring 
atmosphere.

\end{document}